\documentclass[aps,prl,twocolumn,showpacs,10pt]{revtex4-1}
\usepackage{float}
\usepackage{amssymb}
\usepackage{amsmath}
\usepackage{epsfig}
\usepackage{color}
\usepackage{siunitx}
\usepackage[utf8]{inputenc}
\usepackage{graphicx}
\usepackage{enumerate}
\usepackage{array}
\usepackage[normalem]{ulem}
\usepackage{soul}

\usepackage{pdfpages}
\usepackage{pgffor}

\newcolumntype{M}[1]{>{\raggedright}m{#1}}

\newcommand{\beq}{\begin{equation}}
\newcommand{\eeq}{\end{equation}}

\newcommand{\ben}{\begin{eqnarray}}
\newcommand{\een}{\end{eqnarray}}

\newcommand{\Ecoli}{{\it Escherichia coli}}
\newcommand{\ecoli}{{\it E. coli}}
\newcommand{\parS}{{\it parS}}
\newcommand{\parB}{{\it parB}}

\newcommand{\ParABS}{ParAB{\it S}}

\newcommand{\Tw}{\text{Tw}}
\newcommand{\Wr}{\text{Wr}}
\newcommand{\Lk}{\text{Lk}}
\newcommand{\Lko}{\text{Lk}_0}

\newcommand{\mm}{\text{Supp.~Info.}}

\makeatletter
\AtBeginDocument{\let\LS@rot\@undefined}
\makeatother

\begin{document}

\title{Modeling supercoiled DNA interacting with an anchored cluster of proteins:\\
towards a quantitative estimation of chromosomal DNA supercoiling}

\author{J.-C. Walter$^{1*}$, T. Lepage$^2$, J. Dorignac$^1$, F. Geniet$^1$, A. Parmeggiani$^{1,4}$, J. Palmeri$^1$, J.-Y. Bouet$^3$ and I. Junier$^2$}
\email{correspondence should be sent to:\\jean-charles.walter@umontpellier.fr\\ivan.junier@univ-grenoble-alpes.fr\\}
\affiliation{$^1$Laboratoire Charles Coulomb (L2C), Univ. Montpellier, CNRS, Montpellier, France.}
\affiliation{$^2$ CNRS, Univ. Grenoble Alpes, TIMC-IMAG, Grenoble, France}
\affiliation{$^3$ LMGM, CBI, CNRS, Univ. Toulouse, UPS, Toulouse, France.}
\affiliation{$^4$ LPHI, CNRS, Univ. Montpellier, Montpellier, France.}

\begin{abstract}
We investigate the measurement of DNA supercoiling density ($\sigma$) along chromosomes using interaction frequencies between DNA and DNA-anchored clusters of proteins. Specifically, we show how the physics of DNA supercoiling leads, in bacteria, to the quantitative modeling of binding properties of ParB proteins around their centromere-like site, \parS. Using this framework, we provide an upper bound for $\sigma$ in the \Ecoli\ chromosome, consistent with plasmid values, and offer a proof of concept for a high accuracy measurement. To reach these conclusions, we revisit the problem of the formation of ParB clusters. We predict, in particular, that they result from a non-equilibrium, stationary balance between an influx of produced proteins and an outflux of excess proteins, i.e., they behave like liquid-like protein condensates with unconventional ``leaky'' boundaries.
\end{abstract}

\maketitle
	
In most bacteria, DNA is underwound. Despite its critical role for genome structuring~\cite{Vologodskii:1994tj} and coordination of gene expression~\cite{Dorman:2016ky}, measurement of the negative supercoiling along chromosomes remains highly challenging, with both biological and physical difficulties.
Biological difficulties stem from the complex functioning of cells. For instance, a large part of supercoiling is known to be absorbed by various histone-like proteins~\cite{Travers:2005ck}. The remaining supercoiling, which is responsible for the formation of branched plectonemic structures~\cite{Vologodskii1992}, is usually referred to as "free" or "effective"~\cite{Bliska:1987vq}. 
Physical difficulties are inherent in the dual nature of supercoiling. That is, in the absence of topoisomerases, a topologically constrained DNA molecule (such as circular molecules or constrained linear domains~\cite{Liu:1987tq}) is characterized by a constant linking number, $\Lk$, equal to the sum of the twist ($\Tw$), the cumulative helicity of the molecule, and the writhe ($\Wr$), the global intricacy of the molecule~\cite{White:1969fm}. As a consequence, supercoiling, i.e.~the change of $\Lk$ with respect to $\Lko$, the value at rest, leads to changes in the mean values of both $\Tw$ and $\Wr$. Having access to only $\Tw$, when using e.g.~DNA intercaling agents, is thus {\it a priori} insufficient to fully characterize the topological status associated with chromosomal loci~\cite{Lal:2016bv}. This explains why supercoiling density, $\sigma=(\Lk-\Lko)/\Lko$, has been estimated quantitatively using plasmid reporters only, since their compaction level can be quantitatively assessed {\it in vitro} -- see~\cite{Sinden:1980cs} for an exception, although the chromosomal measurement is global, not local. Note, in this regard, that a genetic recombination-based system sensitive to the tightness of plectonemes has been developed to probe supercoiling density along the chromosome~\cite{Booker:2010de,Rovinskiy:2019kx}. The quantitative estimation of $\sigma$ {\it in vivo} still remains problematic, because the method can only be calibrated {\it in vitro}~\cite{Booker:2010de}.
\begin{figure}[t]
\center
\center\includegraphics[width=\linewidth]{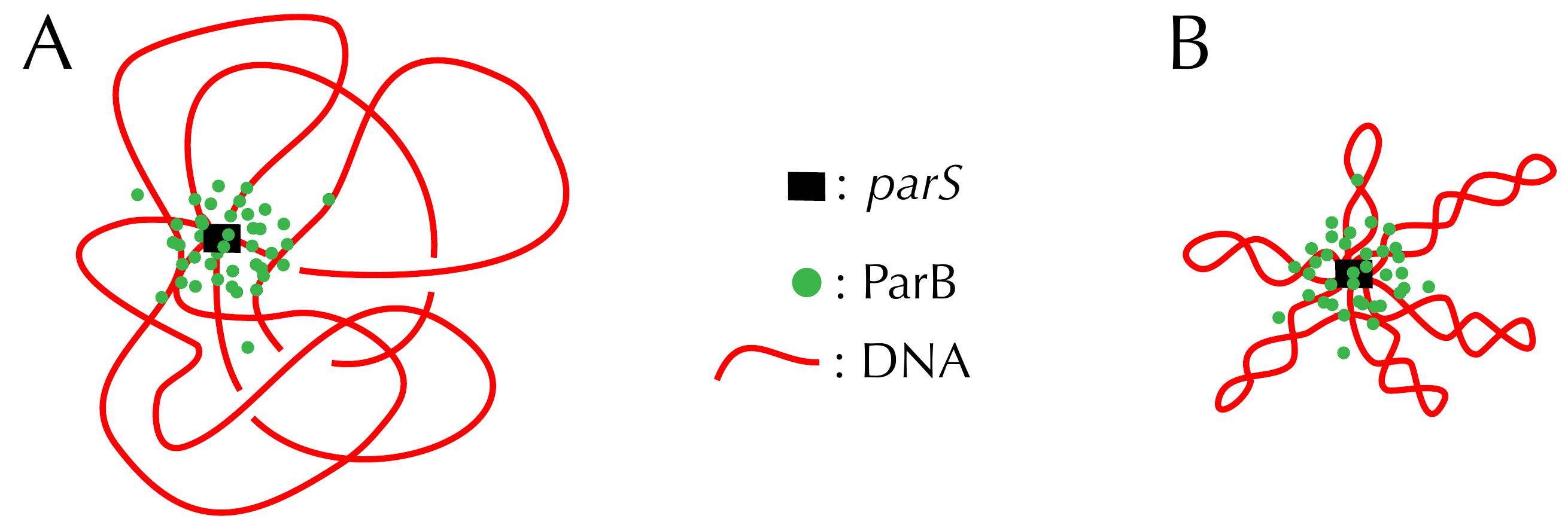}
\caption{{\it Stochastic binding model}. When DNA enters the high concentration region of the \parS-anchored cluster of ParB, crosslinking with ParB occurs with high probability during the ChIP-seq protocol. Compared to relaxed DNA (A), supercoiling DNA (B) tends to increase DNA compaction and, hence, crosslinking with DNA loci far from \parS.}
\label{fig:sketch} 
\end{figure}

Here, we investigate the possibility of measuring the effective chromosomal supercoiling density using DNA binding properties of the centromere-binding protein ParB, part of the active \ParABS~system of DNA segregation. Specifically, it has been argued that the capture by chromatin immuno-precipitation sequencing (ChIP-seq) of the binding of ParB onto DNA in the vicinity of its specific binding site (\parS) is driven by stochastic binding involving DNA looping properties~\cite{Sanchez15} (Fig.~\ref{fig:sketch}).
More precisely, ParB proteins cluster around \parS~\cite{Sanchez15,Debaugny:2018iz} through a phase separation-like mechanism~\cite{David2018}. In this context, it has been shown that only a process of looping, which brings DNA loci inside the cluster, can explain the long range decay of the ParB binding profile as the genomic distance to \parS~increases (black curves in Fig.~\ref{fig:plasm})~\cite{Sanchez15,Walter18}.
Knowing that supercoiling properties strongly influence DNA looping properties, we thus assess whether a quantitative reproduction of the ParB binding profile in the vicinity of \parS\ is possible using a model {\it with no other free parameter than DNA supercoiling density ($\sigma$)}. To this end, we consider, on one hand, a realistic model of supercoiled DNA that has been independently calibrated using single-molecule techniques and, on the other hand, an independent estimation of the size of the ParB cluster using high-resolution microscopic experiments.

Compared to the previous stochastic binding model where a very small DNA persistence length ($\SI{10}{bp}$), difficult to justify on physical grounds, was required to match experimental data~\cite{Sanchez15}, here we show, using numerical simulations of realistic long (i.e.~$\geq \SI{30}{kb}$) molecules, that DNA supercoiling indeed leads to a quantitative reproduction of ChIP-seq ParB binding profiles. In this context, we provide a bound for the chromosomal supercoiling density, propose new experimental protocols to further fix its exact values and demonstrate, for the first time to our knowledge, the consistency between chromosomal and plasmid measurements. In addition, we provide novel insights into the physical properties of ParB clusters. In particular, we predict a cluster shape that differs from the usual sharp boundaries of liquid droplets.
Namely, we show that the cluster density profile display unconventional ``leaky'' boundaries, which can be explained as a perturbation induced by a source of proteins located at the edge of the cluster core. Altogether, our work thus offers insights into both bacterial DNA organization and liquid-like protein condensates. It also offers a proof of concept for measuring chromosomal supercoiling with high accuracy.

\paragraph{{\bf Stochastic Binding model.}}

ChIP-seq detection of DNA-bound proteins involves sub-nm crosslinking between DNA and proteins~\cite{Hoffman:2015gi}. Thus we expect that the non-specific ParB binding profile results from "collisions" between DNA and the ParB proteins located in the \parS-anchored cluster (Fig.~\ref{fig:sketch}). We therefore suppose that, except at \parS, the timescale for ParB to unbind DNA is much shorter than the timescale for DNA to diffuse away from the location where binding occurs (instantaneous unbinding hypothesis). The modeled non-specific ParB binding profile, $B(s)$, thus reads~\cite{Sanchez15}:
\beq
B(s)=\int 4\pi r^2 P_s(r) C(r) dr\,.
\label{eq:profile}
\eeq
$P_s(r)$ describes DNA looping properties: it stands for the equilibrium probability distribution function for a DNA locus at a genomic distance $s$ from \parS~ to be located at a distance $r$ from \parS\ in the three-dimensional space.
For simplicity, here we neglect effects coming from the interaction between DNA and the cluster, therefore $P_s(r)$ is computed by considering an isolated DNA chain. 

$C(r)$ stands for the probability to find a ParB protein at distance $r$ from \parS. Although its exact shape is not known (see below for predictions), we have $C(r=0)=1$ by definition of the strong binding of ParB to \parS. Next, the full width at half maximum of the cluster, $\omega$, has been estimated using high-resolution fluorescent microscopy, leading to $\omega_{exp}=37 \pm \SI{5}{nm}$~\cite{Guilhas:2020jc}. In this experiment, cell contents were chemically fixed so that $\omega_{exp}$ refers to the probability $C^{(0)}(x)$ to find a ParB protein at a distance $x$ from the cluster center, not from \parS, with $C_{exp}^{(0)}(\omega_{exp}/2)=0.5$. Considering the positional degrees of freedom of the cluster center with respect to \parS, we then have $C(r)=\int_{0}^{\infty} dx\; \Pi_r(x) C^{(0)}(x)$, where $\Pi_r(x)$ stands for the probability density of finding the cluster center at a distance $x$ given a point at distance $r$ from \parS~(\mm). 

\paragraph{{\bf Self-avoiding rod-like chain model of DNA.}}

We consider a realistic $\SI{30}{bp}$ resolution polymer model of bacterial DNA, namely the self-avoiding rod-like chain (sRLC) model~\cite{Vologodskii1992} (detailed simulation procedure in~\cite{Lepage2017}). Specifically, DNA is modeled as a discrete chain of $\SI{10.2}{nm}$ long ($\SI{30}{bp}$ of B-DNA) articulated hard-core cylinders, with radius $r_e=\SI{2}{nm}$ reflecting the short-range electrostatic repulsions of DNA for {\it in vivo} salt conditions~\cite{Lepage2015}. The chain is iteratively deformed using crankshaft elementary motions with Metropolis-Hastings transition rates, under the condition that it does not cross itself. Each articulating site is associated with bending and torsional energies such that the resulting persistence length ($\SI{50}{nm}$ or, equivalently, $\SI{147}{bp}$) and torsional length ($\SI{86}{nm}$) are typical of B-DNA for {\it in vivo} salt conditions~\cite{Vologodskii1992,Strick:2003dp,Lepage2015}.

Here, we discuss results obtained with a $\SI{30}{kb}$ long chain by varying $\sigma$ from $0$ to $-0.08$ slowly enough so that chain statistical properties are insensitive to the associated rate of change (see Fig.~S1 and simulation details in \mm). Simulated conformations are thus expected to reflect thermodynamic equilibrium, even at low values of $\sigma$ where plectonemes are tight. We further checked that our results did not depend significantly on the length of the chain by performing additional simulations of $\SI{60}{kb}$ long chains (Fig.~S2). Note, here, that the motivation to work with $\sigma \geq -0.08$ is both biological and physical: in the worst case of topoisomerase mutants, the total supercoiling density in \ecoli\ has been shown to remain above $-0.08$~\cite{Bliska:1987vq}, while recent work has revealed the existence of a transition toward a hyperbranched regime occurring at $\sigma \simeq -0.08$~\cite{Krajina2016}, which is beyond the scope of our discussion.

\begin{figure}[t]
\center
\center\includegraphics[width=\linewidth]{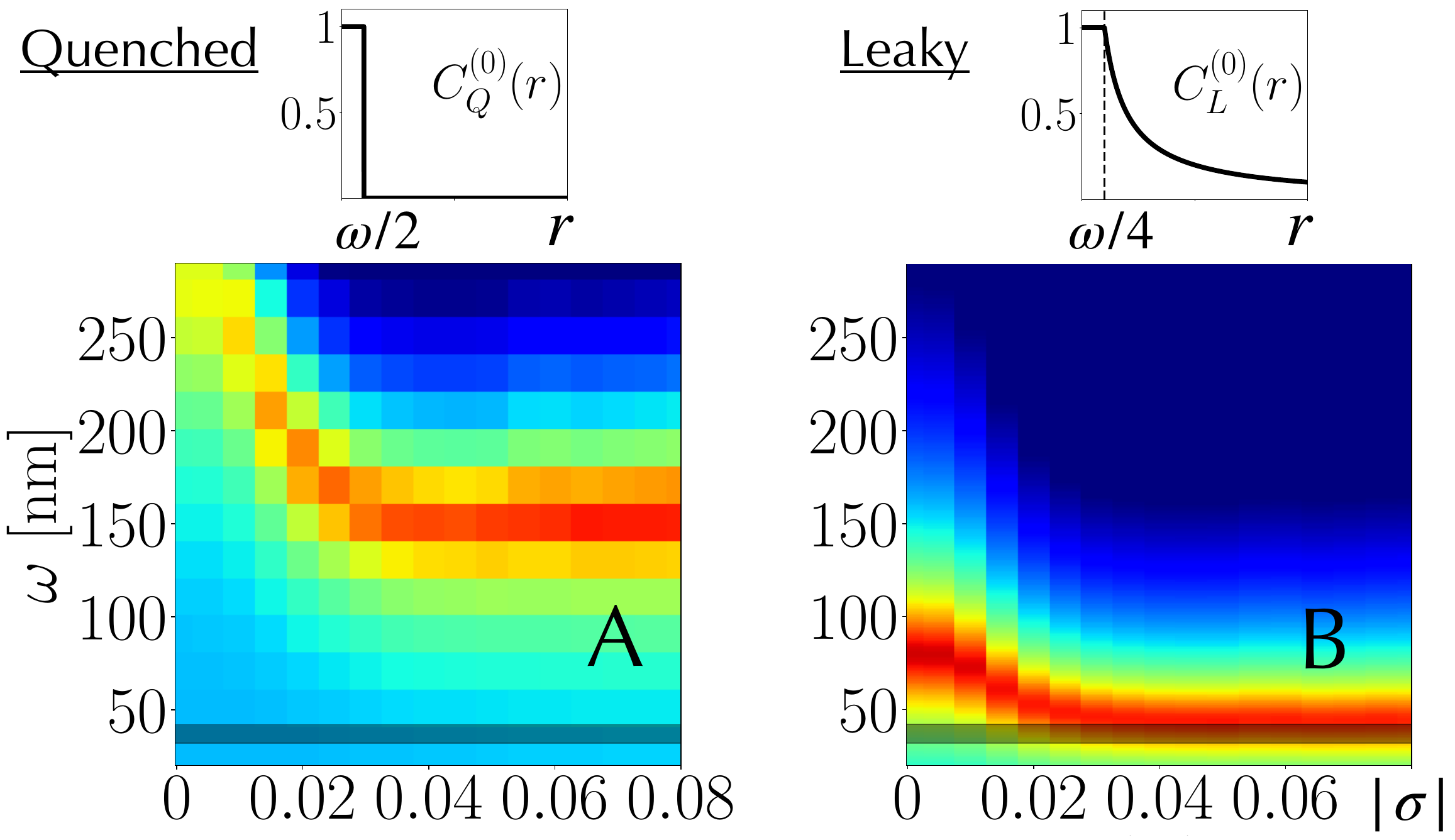}
\caption{{\it Capturing chromosomal binding profiles}. Root mean squared deviation between modeled binding profiles and ChIP-seq chromosomal data (curves can be found in Fig.~S3); the redder the pixel, the smaller the deviation (arbitrary scale). The horizontal dark band indicates $\omega_{exp}(\SI{37}{nm} \pm \SI{5}{nm})$. (A) The best models with quenched clusters imply a large cluster with $\omega_{best}=\SI{150}{nm}$. (B) In contrast, the best models with leaky clusters imply cluster sizes very close to microscopic data when $\sigma \lesssim -0.04$. In this regime, all best models indeed correspond to $\omega_{best}=\SI{44}{nm}$.}
\label{fig:chrom} 
\end{figure}

\paragraph{{\bf Leaky vs quenched cluster.}}
Having in hand the corresponding $P_s(r)$ for $\sigma \in [-0.08,0]$, we now consider the spatial distribution of ParB proteins associated with the \parS-anchored clusters. In this regard, high-resolution microscopic measurements~\cite{Guilhas:2020jc} suggest that these clusters result from a phase transition-like mechanism. Theoretical models further suggest that this phase transition is unconventional as it implies a framework of a lattice gas on a fluctuating polymer~\cite{David2018}. Moreover, the physical formation of a cluster is likely to interfere with biological processes like, e.g., the production of ParB close to the cluster, just as membrane proteins are often produced close to the membrane~\cite{Libby:2012js}. In other words, the spatial distribution of ParB proteins around \parS~ remains an open question.

Here, we investigate two extreme cases for the shape of these clusters, referred to as {\it quenched} and {\it leaky}. A quenched cluster (Fig.~\ref{fig:chrom}A) is defined by $C_Q^{(0)}(r)=\theta(\omega/2-r)$, with $\theta$ the Heaviside function. It corresponds to the conventional sharp interface of a droplet. A leaky cluster (Fig.~\ref{fig:chrom}B) further includes the stationary solution of a diffusion process where ParB proteins are continuously produced at the edge of the cluster core and diluted due to cell growth and division (\mm). That is, the leaky cluster releases proteins in excess, while $C_{L}^{(0)}=1$ for $r\leq \frac{\omega}{4}$ (cluster core) reflects the saturation regime in which experiments are performed~\cite{Debaugny:2018iz}. As a result, $C_L^{(0)}$ includes a $1/r$ long range decay such that $C_L^{(0)}(r)=\theta(\frac{\omega}{4}-r)+\frac{\omega}{4r}\theta(r-\frac{\omega}{4})$. Note that for both quenched and leaky cases, the full width at half maximum of $C^{(0)}$ is equal to $\omega$.

We computed binding profiles for $\sigma$ ranging in $[-0.08,0]$ and for values of $\omega$ between $\SI{10}{nm}$ and $\SI{300}{nm}$. We compared them with profiles obtained for \ecoli\ by inserting \parS\ along the chromosome (black curve in Fig.~S3) -- only one side of the chromosome is analyzed as the other side is distorted by the presence of strong promoter regions~\cite{Debaugny:2018iz}. In this experiment, $10$ \parS\ sites interspersed by $43$ base pairs, as found in the natural \parS\ region, were inserted at {\it xylE} locus~\cite{Debaugny:2018iz}. A careful analysis of the binding properties among these \parS\ sites actually revealed significant variations of the ChIP-seq signal, which was thus normalized with respect to the maximum value. The origin of the curvilinear abscissa $s$ was set right at the edge of the most extreme \parS\ site.

We are interested in explaining the global shape of the binding profile as it is expected to reflect generic polymer physics principles. To that end, we quantify the explanatory power of each model by reporting the root mean square deviation with respect to the experimental binding profile for $s \in [\SI{1.5}{kb},\SI{9}{kb}]$. Both the lower and upper bounds at $\SI{1.5}{kb}$ and $\SI{9}{kb}$, respectively, are used to avoid specific, reproducible  distortions of the signal associated with the presence of gene promoters and sites for regulatory DNA proteins~\cite{Debaugny:2018iz}.

\begin{figure}[t]
\center
\center\includegraphics[width=\linewidth]{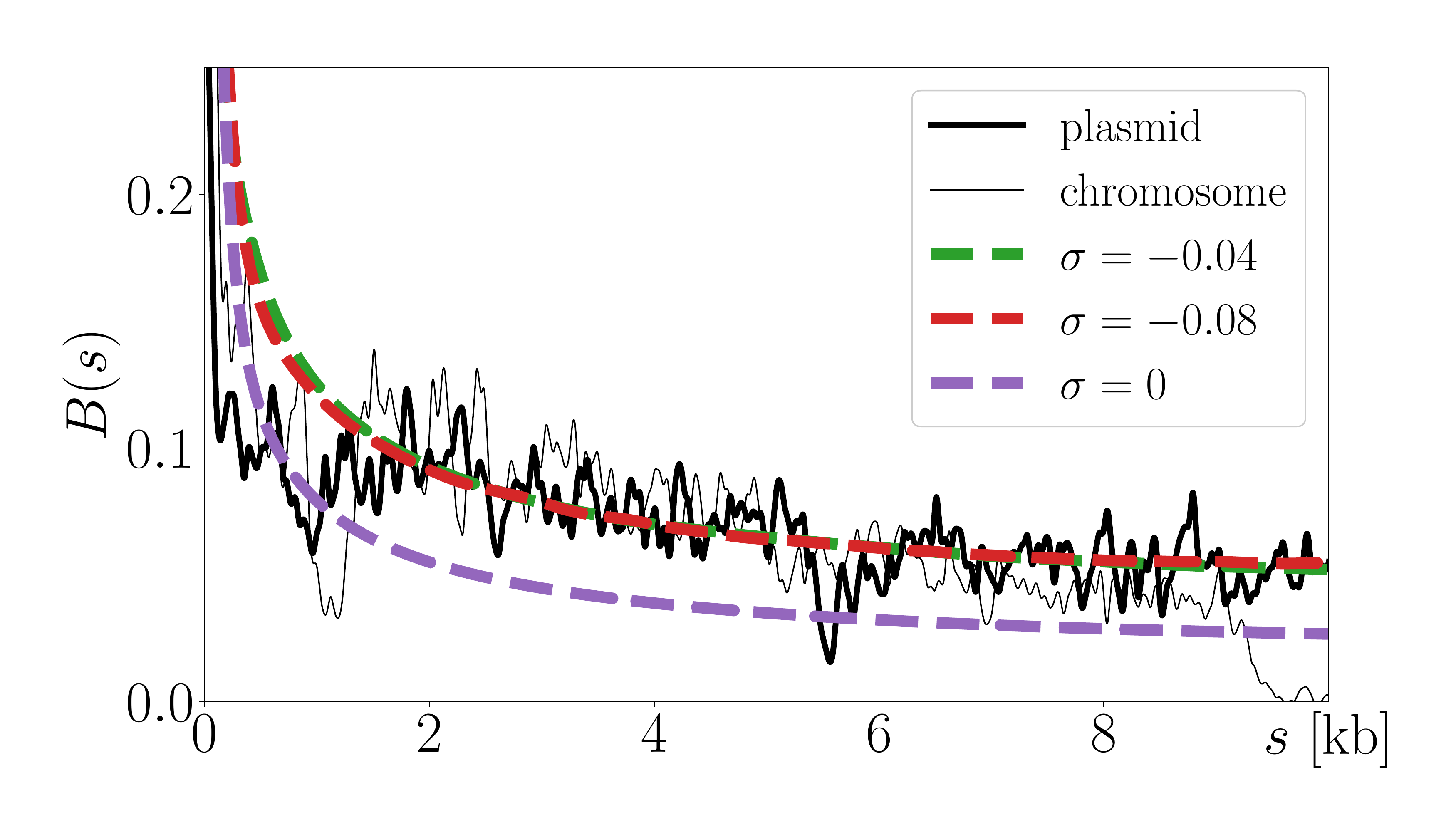}
\caption{Compared to chromosome data (thin black curve), the leaky models with $\omega_{best}=\SI{43}{nm}$ (smooth curves) and $\sigma\leq-0.04$ (green and red curves) capture even better plasmid data (thick black curve).
We notice that model predictions for $\sigma=-0.04$ and $\sigma=-0.08$ are almost undistinguishable.
}
\label{fig:plasm} 
\end{figure}

We find that both quenched and leaky clusters can accurately capture experimental data (Fig.~S3). However, the best quenched models are found at $\omega_{best}=\SI{150}{nm}$ (Fig.~\ref{fig:chrom}A), which is much larger than $\omega_{exp}$. In contrast, the best leaky models are found at $\omega_{best}=\SI{44}{nm}$ when $\sigma \lesssim -0.04$ (Fig.~\ref{fig:chrom}B). That is, they explain data in the physiologically relevant plectonemic regime of bacterial DNA. They also solve the small DNA persistence length issue associated with the previous version of the stochastic binding model where DNA supercoiling was neglected~\cite{Sanchez15} -- a small persistence length was indeed needed to ``mimic'' compaction due to plectonemes.
Interestingly, compared to chromosomal \parS\ data, ParB binding profiles in the vicinity of a \parS\ located on a plasmid ($\SI{100}{kb}$ long F-plasmid~\cite{Debaugny:2018iz}) show less distortion (Fig.~\ref{fig:plasm}) -- just as for the chromosome, only one side of the plasmid is analyzed as the other side is distorted by binding sites for a replication initiator~\cite{Sanchez15}. In this context, the best leaky models lead to similar model parameters ($\omega_{best}=\SI{43}{nm}$ when $\sigma \lesssim -0.04$), while providing an even better match with the data (Fig.~\ref{fig:plasm}). Compared to the chromosomal situation where the gene \parB\ is located $\SI{750}{kb}$ away from \parS, this better match might reflect a phenomenology of the plasmid fitting particularly well the leaky situation, with \parB\ located only $\SI{74}{bp}$ away from \parS~\cite{Bouet:2006bp}. The hypothesis of a source located on the edge of the cluster core is indeed even more relevant since the production of proteins in bacteria often occurs close to their gene~\cite{Llopis:2010bm}.

\begin{figure}[t]
\centering
\includegraphics[width=\linewidth]{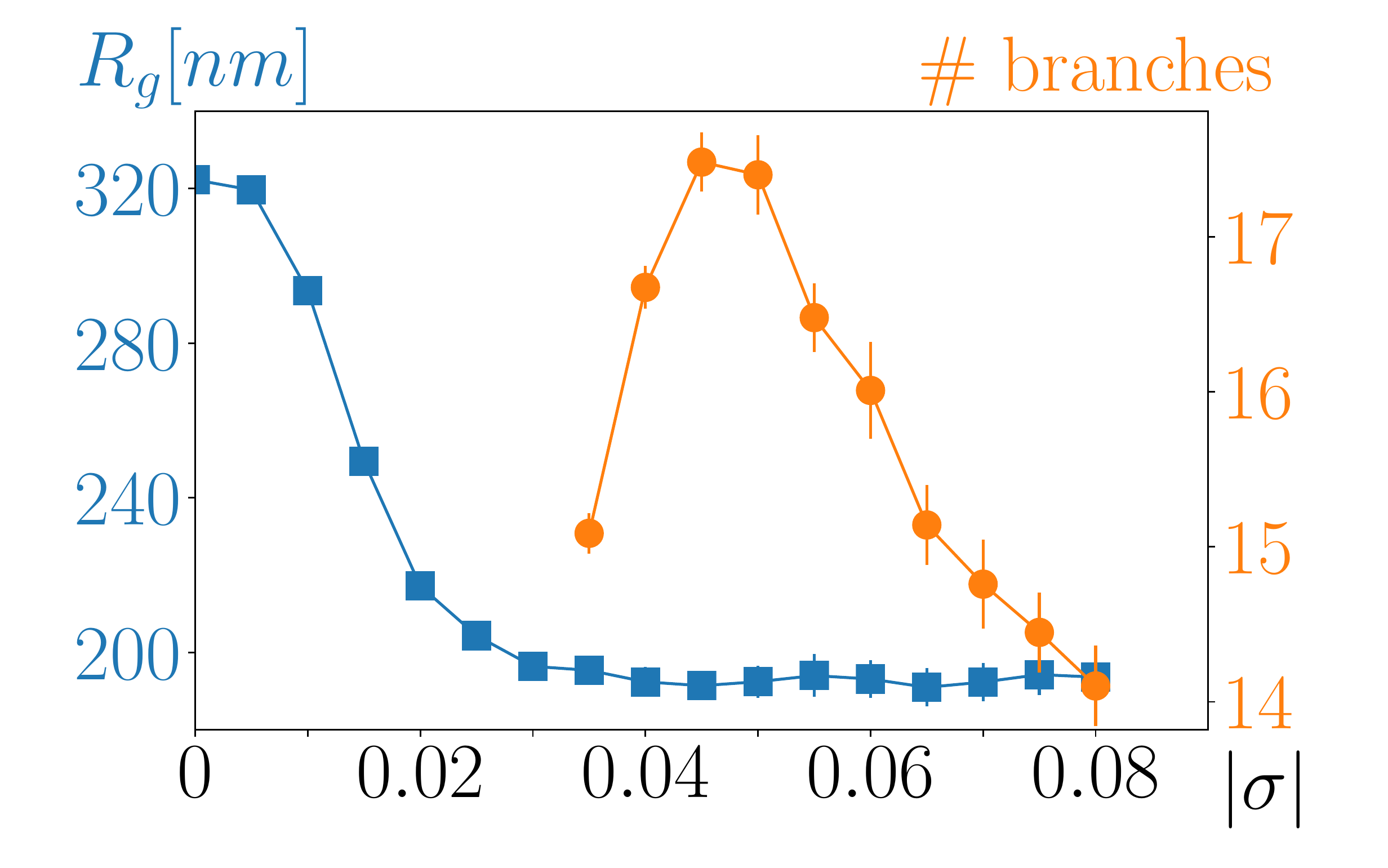}
\caption{The radius of gyration of a \SI{30}{kb} long circular molecule plateaus at $|\sigma|\simeq 0.04$. The number of plectonemic branches is non-monotonous, reaching a maximum at $|\sigma|\simeq 0.05$. The error bars correspond to the standard error of the mean computed over 20 different simulation runs (see \mm~for details).}
\label{fig:Rg_branches}
\end{figure}

\paragraph{{\bf $\sigma$-sensitive probes for strong supercoiling.}}

While leaky models with experimentally relevant $\omega$ capture experimental data rather well, resulting binding profiles are almost indistinguishable for $\sigma \in [-0.08,-0.04]$ (Fig.~\ref{fig:plasm}). This lack of sensitivity is concomitant with a poor variation of the radius of gyration (blue curve in Fig.~\ref{fig:Rg_branches}) in the plectonemic regime. Note that, in contrast, branching properties can vary significantly in this regime~\cite{Vologodskii:1994tj,Krajina2016}. For instance, we find that the number of plectonemic branches reaches a maximum at $\sigma \simeq -0.05$ (orange curve in Fig.~\ref{fig:Rg_branches}), in accord with previous analyses with smaller molecules~\cite{Vologodskii:1994tj} and with a minimum value of the hydrodynamic radius for $\SI{10}{kb}$ long plasmids~\cite{Wang:1974jm,Vologodskii:1994tj,Krajina2016}.

\begin{figure}[t]
\center
\center\includegraphics[width=\linewidth]{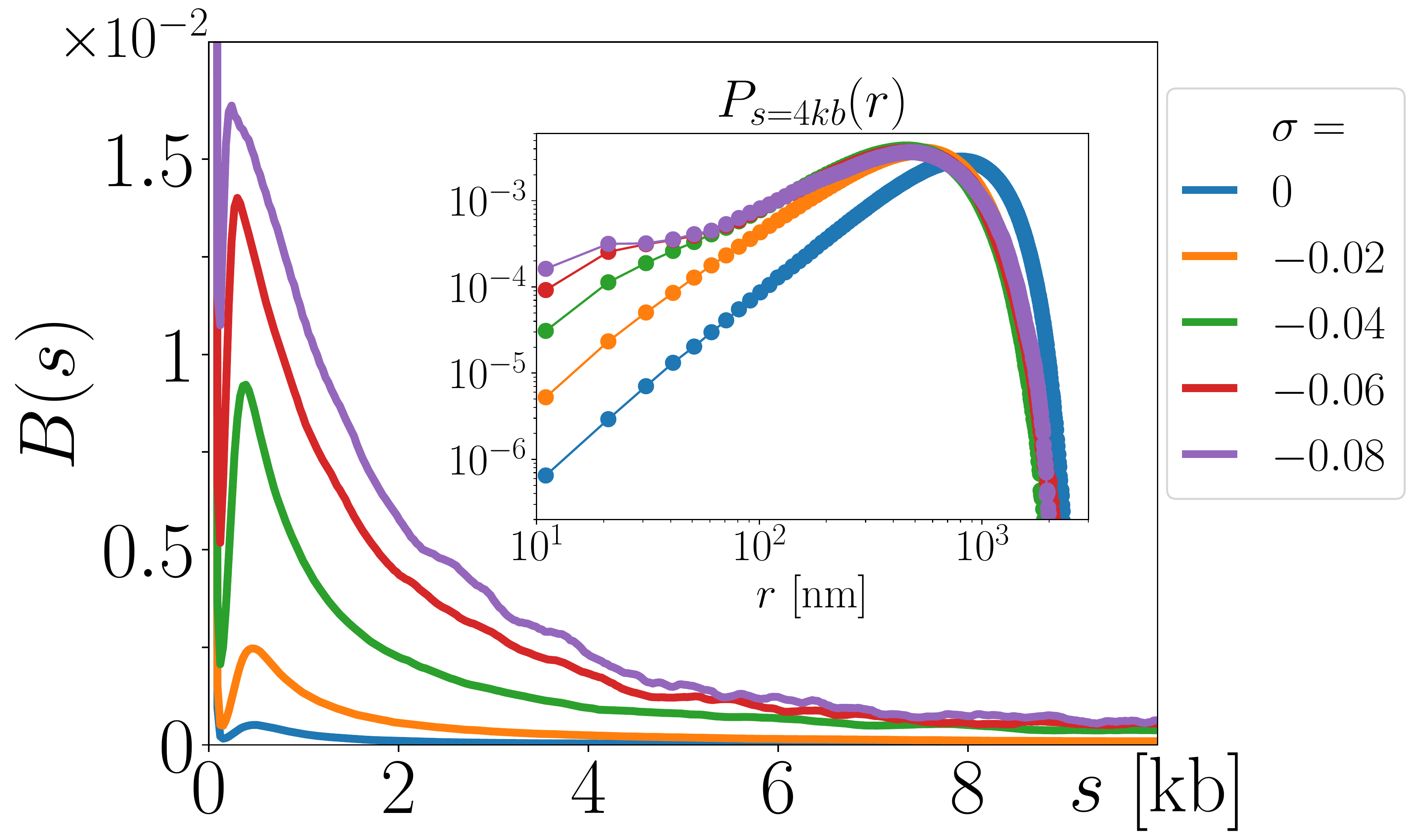}
\caption{{\it A $\sigma$-sensitive probe.}
With a small quenched cluster ($\omega=\SI{20}{nm}$), binding profiles are well separated for values of $\sigma \in [-0.08,0]$ differing by $0.01$, which would thus provide a reasonable precision for measuring supercoiling. Inset: in the plectonemic regime ($\sigma \leq -0.04$), the spatial distribution of distances between loci differ significantly only at small distances associated with plectonemic intertwining properties.}
\label{fig:discrim} 
\end{figure}

A natural question, then, is whether it is possible to build a probe that is sensitive to variations of $\sigma$ for strong supercoiling. Interestingly, we have found a possible solution consisting of a system that senses intertwining properties of plectonemes, in the spirit of the $\gamma\delta$ recombination system~\cite{Booker:2010de}. In that respect, one would need a quenched (instead of a leaky) cluster that is small enough such that the binding properties of proteins is sensitive to the diameter and pitch of plectonemes~\cite{Marko1995,Barde2018}. For instance, our simulations reveal a strong sensitivity of $P_s(r)$, at the kb genomic scale for $s$,  with respect to {\it all values of $\sigma$} for spatial distances $r$ on the order of $\SI{10}{nm}$ (inset of Fig.~\ref{fig:discrim}). One can verify, then, that a quenched cluster with $\omega=\SI{20}{nm}$ provides well-distinct binding profiles for $\sigma \in [-0.08,0]$ (Fig.~\ref{fig:discrim}). Notice the much smaller values of $B(s)$ in this case, compared e.g.~to results in Fig.~\ref{fig:plasm}. ParB ChIP-seq experiments can nevertheless report very low binding frequencies as demonstrated by titration assays~\cite{Debaugny:2018iz}.

\paragraph{{\bf Discussion and perspectives.}}

We have shown that the binding profile of ParB proteins in the vicinity of \parS\ can be quantitatively explained considering a stochastic binding process between a supercoiled DNA and proteins that are issued from a saturated \parS-anchored core cluster. To this end, we had to consider clusters from a non-equilibrium, stationary perspective, with the presence of a spatially localized source and sink. Biologically, the sink reflects protein dilution due to cell growth and division, while the source may arise from two effects: the continuous activity of genes producing new proteins in a saturated cluster and the effect of an unconventional liquid-like nature of the cluster.
Namely, we predict the cluster core to result from a balance between an influx of continuously produced proteins and an outflux of proteins in excess. In the plasmid, the situation may even be more prototypical with the production of ParB occurring close to \parS.

In this context, and for the first time to the best of our knowledge, we provide an upper bound ($\sigma\approx-0.04$) for the {\it in vivo} supercoiling density at a chromosomal location of a bacterium (\ecoli\ during its exponential growth) and we show that it also holds for plasmids. Interestingly, this value corresponds to the onset of the plectonemic regime characterized by a poor variation of the radius of gyration, on one hand, and a significant variation of branching properties, on the other hand. Importantly, we also offer a proof of concept to obtain a finer estimate of the supercoiling density. Specifically, in the spirit of existing genetic recombination-based probes~\cite{Booker:2010de,Rovinskiy:2019kx}, we demonstrate that a small quenched cluster provides a supercoiling-sensitive probe as it "senses" physical properties of plectonemes.

Compared to "biological" genetic recombination-based probes, our "physical" ChIP-seq-based probe is expected to be much less invasive. It should also be less sensitive to molecular environment as it is based on generic (polymer) physics properties -- for instance, recombination-based systems depend on (slow) enzymatic recombinase reactions, whose quantitative modeling has, to the best of our knowledge, remained elusive. In practice, while genetic design of quenched clusters of ParB proteins might be tricky, transcription factors could provide an efficient system. These proteins have indeed the capacity of binding both cognate DNA sites strongly and other DNA sites non-specifically with (short) millisecond residence times~\cite{Elf:2007um}. They could also be used in conjunction with a  DNA methyltransferase to generate methylation (instead of binding) profiles without the need of crosslinking stages~\cite{Redolfi:il}. Finally, a sensitive system would require having the designed artificial DNA devoid as much as possible of interfering biological elements, such as gene promoters, which distort the utilizable physical signal. Along this line, one would like to have an explicit description of ParB nucleation and diffusion properties to develop a detailed model of the interactions between ParB and DNA using e.g.~molecular dynamics approaches. In particular, the discrepancy between experimental and modeling profiles below $\sim\SI{1}{kb}$ (Fig.~\ref{fig:plasm}) might be the result of our approximation of neglecting hard-core interactions between ParB proteins and DNA. At large scales, cellular confinement of DNA should also be included in the model. We note, nevertheless, that a complete picture would require studying the melting of a plectonemic tree-like structure at the chromosome scale, which is currently beyond the capacities of numerical simulations.

\begin{acknowledgements}
We thank Daniel Jost for useful suggestions. JCW was supported by a ``Modélisation pour le Vivant'' CNRS Grant (CoilChrom). J.Y.B. is supported by an AO80Prime (Numacoiled) CNRS grant. I.J. was supported by an ATIP-Avenir grant (Centre National de la Recherche Scientifique).
\end{acknowledgements}


%



\section*{Supplementary material}

# Supplementary information:
# Modeling supercoiled DNA interacting with an anchored cluster of proteins: towards a quantitative estimation of chromosomal DNA supercoiling

J.-C. Walter, T. Lepage, J. Dorignac, F. Geniet, A. Parmeggiani, J. Palmeri, J.-Y. Bouet and I. Junier

## CONTENTS

## I. THE LEAKY CLUSTER MODEL

For the quenched cluster model, all ParB proteins are confined to a spherical volume with radius $\omega/2$, what we refer to as a "cluster core" in the following. For the leaky cluster model, we consider an additional process where ParB proteins are produced at the edge of a cluster core and diffuse in the volume – we remind that the leaky core cluster radius is smaller than $\omega/2$; it is set to obtain the same $\omega$, i.e. the same full width at half maximum of $C^{(0)}$. This production process accounts for the fact that, for the experimental conditions we consider, ParB proteins are produced in excess with respect to the capacity of the cluster core. The location of the source at the edge of the cluster may then model several situations (see main text for further details and references): either ParB production occurs close to the cluster as in the case of the plasmid or of membrane proteins that are often produced close to the membrane; or the cluster continuously "radiates" ParB proteins, which would be in accord with the prediction that ParB-ParB interactions are on the order of $k_B T$ inside the cluster. Since diffusing proteins are continuously diluted due to cell growth and cell division, we further consider an annihilation process. In this context, the *a priori* time-dependent quantity $C^{(0)}(r, t)$, which is itself proportional to the concentration of proteins, is governed by the following equation of diffusion (in spherical coordinates):

$$\partial_t C^{(0)} - \frac{D}{r^2} \partial_r \left[ r^2 \partial_r C^{(0)} \right] = S(r) - \Gamma(r) C^{(0)} \tag{1}$$

where $S(r)$ is the time-independent protein source contribution and $\Gamma(r)$ is a time-independent dilution rate associated with the annihilation process.

Here, we consider a source located at $r = \rho$ (radius of the cluster core) and look for the solution when $r > \rho$, that is, where $S(r) = 0$. We also consider the biologically relevant hypothesis that dilution occurs in principle far from the cluster; in practice we consider that dilution occurs at infinity so that $\Gamma(r) = 0$ as well. That is, the source and annihilation processes are boundary conditions that further specify the homogeneous solution of the problem. Finally, we are interested in the stationary solution, $C^{(0)}(r)$, such that $\partial_t C^{(0)} = 0$. We are thus left with a simple homogeneous equation:

$$\frac{D}{r^2} \partial_r \left[ r^2 \partial_r C^{(0)} \right] = 0 \tag{2}$$

whose solution reads:

$$C^{(0)}(r) = A + \frac{B}{r} \tag{3}$$

In principle, $A$ and $B$ are determined by the boundary conditions related to the production process at the edge of the cluster and to the annihilation process at infinity. Here, we consider that there is saturation at the edge of the cluster ($C^{(0)}(\rho) = 1$) and that the concentration far from the cluster can be neglected ($C^{(0)}(\infty) = 0$), leading to $A = 0$ and $B = \rho$. Altogether, the leaky cluster model is thus defined by:

$$C^{(0)}(r) = \begin{cases} 1 & \text{if } r \leq \rho \\ \frac{\rho}{r} & \text{if } r > \rho \end{cases} \tag{4}$$

$$\rho = \frac{\omega}{4} \tag{5}$$

## II. COMPUTING $C(r)$ FROM THE KNOWLEDGE OF $C^{(0)}(x)$

Here, we aim at computing the probability $C(r)$ to find a ParB protein at a point $P$ located at a distance $r$ of *parS*, knowing that the center of the cluster can be found anywhere in the *parS*-centered sphere with radius $\rho$ (dashed red circle in Fig. 1), with $\rho = \omega/2$ and $\rho = \omega/4$ for the cases of the quenched and leaky clusters, respectively (see above and main text for the definition of $\omega$). Specifically, given the probability $\Pi_r(x)$ to find the cluster center at a distance $x$ of $P$, knowing that the latter is located at distance $r$ of *parS*, and given the probability $C^{(0)}(x)$ to find a ParB protein at a distance $x$ of the cluster center, $C(r)$ can be written as:

$$C(r) = \int_0^\infty dx\, \Pi_r(x) C^{(0)}(x), \tag{6}$$

We next consider the general situation where $C^{(0)}(r) = \theta(\rho - r) + \theta(r - \rho) \times f(r)$ where: i) $\rho = \omega/2$ and $f(r) = 0$ for the quenched cluster and ii) $\rho = \omega/4$ and $f(r) = \rho/r$ for the leaky cluster. Considering separately the cases $r \leq \rho$ and $r \geq \rho$, one can then show that $C(r)$ reads:

$$C(r) = \theta(\rho - r)\left[\int_0^{\rho-r} dx\, \Pi_r^{(1)}(x) C^{(0)}(x) + \int_{\rho-r}^{\rho+r} dx\, \Pi_r^{(2)}(x) C^{(0)}(x)\right] + \theta(r - \rho)\int_{\rho-r}^{\rho+r} \Pi_r^{(2)}(x) C^{(0)}(x) \tag{7}$$

$\Pi_r^{(1)}(x) = 3x^2/\rho^3$ (which is independent of $r$) stands for the probability density to pick a point on the $P$-centered sphere with radius $x$ given a random process of picking points uniformly in the *parS*-centered spherical volume with radius $\rho$, knowing that the former is entirely included in the latter, i.e. knowing that $r + x \leq \rho$ (Fig. 1A). $\Pi_r^{(2)}(x) = \frac{3x}{4r\rho^3}\left(\rho^2 - (r-x)^2\right)$ is the corresponding probability in the situation where either $r \leq \rho$ and $r + x \geq \rho$

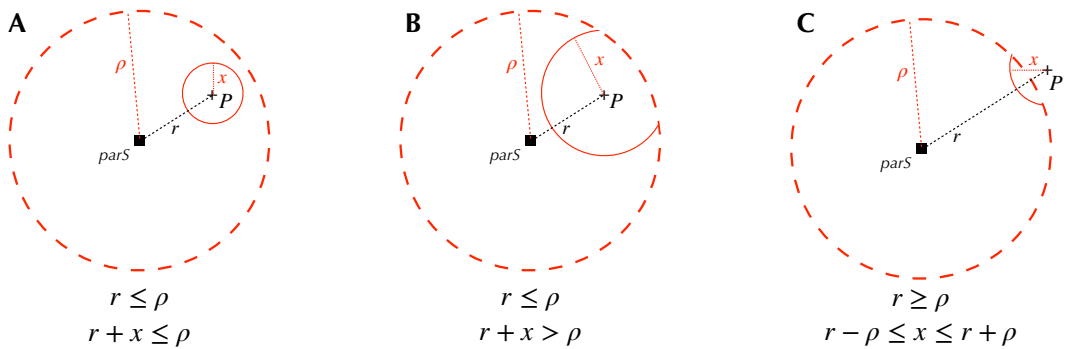

FIG. 1. *The different cases to consider to compute $C(r)$ as a function of $C^{(o)}(x)$. $P$ indicates a point at distance $r$ from parS at which we compute $C(r)$. The small red circle and arcs of a circle indicate possible locations of the center of the cluster knowing it is located at a distance $x$ from $P$ (and, hence, contributing by $C^{(o)}(x)$). The large dashed red circle indicates the maximal distance between parS and the center of the cluster core. A) The distance $r$ and $x$ are such that all the positions on the $P$-centered sphere of radius $x$ are possible for the center of the cluster core, leading to $\Pi_r(x) = \Pi_r^{(1)}(x) = 3x^2/\rho^3$. B) $P$ is located inside the volume accessible by the cluster core but $x$ is large enough such that only part of the $P$-centered sphere of radius $x$ contributes to the signal, leading to $\Pi_r(x) = \Pi_r^{(2)}(x) = \frac{3x}{4r\rho^3}\left(\rho^2 - (r-x)^2\right)$. C) $P$ is located outside the volume accessible by the cluster core such that, just as in B, $\Pi_r(x) = \Pi_r^{(2)}(x)$.*

(Fig. 1B) or $r \geq \rho$ and $r - \rho \leq x \leq r + \rho$ (Fig. 1C), i.e. when the $P$-centered sphere intersects only partially with the *parS*-centered spherical volume.

Interestingly, one can check that considering $C_0(r)$ as an approximation of $C(r)$ leads to similar results, with significant differences only for small binding probabilities. As a consequence, results presented in Figs. 2 and 3 of the main text are qualitatively similar when using $C^{(0)}(r)$ in place of $C(r)$, with in particular the same values of best parameters. Results in Fig. 4 are also qualitatively similar. The overall values of $B(s)$ are nevertheless approximately two fold smaller when using $C^{(0)}(r)$ in place of $C(r)$, i.e. the positional degrees of freedom tend to increase the overall contact frequency between DNA and the core cluster, as expected.

### A. Quenched cluster: $C_Q^{(0)}(x) = \theta(\rho - x)$

In this case, using the notation $W_{r_1}^{r_2}(r) = \theta(r_2 - r) \times \theta(r - r_1)$ for the window function, we find:

$$C_Q(r) = W_0^\rho(r) \left[ \left( \frac{\rho - r}{r} \right)^3 + P_r^{(2)}(\rho) - P_r^{(2)}(\rho - r) \right] + W_\rho^{2\rho}(r) \left[ P_r^{(2)}(\rho) - P_r^{(2)}(r - \rho) \right] \tag{8}$$

$$P_r^{(2)}(x) = \int^x \Pi_r^{(2)}(y) dy = \frac{3}{8} \frac{x^2}{r^2} \left( \frac{r}{\rho} - \left( \frac{r}{\rho} \right)^3 \right) + \frac{1}{2} \frac{x^3}{\rho^3} - \frac{3}{16} \frac{x^4}{r \rho^3} \tag{9}$$

$$\rho = \omega/2 \tag{10}$$

### B. Leaky cluster: $C_L^{(0)}(x) = \theta(\rho - x) + \theta(x - \rho) \times \rho/x$

In this case, we find:

$$C_L(r) = W_0^\rho(r) \left[ \left( \frac{\rho - r}{r} \right)^3 + P_r^{(2)}(\rho) - P_r^{(2)}(\rho - r) + Q_r^{(2)}(r + \rho) - Q_r^{(2)}(\rho) \right] +$$
$$W_\rho^{2\rho}(r) \left[ P_r^{(2)}(\rho) - P_r^{(2)}(r - \rho) + Q_r^{(2)}(r + \rho) - Q_r^{(2)}(\rho) \right] +$$
$$W_{2\rho}^\infty(r) \left[ Q_r^{(2)}(r + \rho) - Q_r^{(2)}(r - \rho) \right] \tag{11}$$

$$P_r^{(2)}(x) = \int^x \Pi_r^{(2)}(y) dy = \frac{3}{8} \frac{x^2}{r^2} \left( \frac{r}{\rho} - \left( \frac{r}{\rho} \right)^3 \right) + \frac{1}{2} \frac{x^3}{\rho^3} - \frac{3}{16} \frac{x^4}{r \rho^3} \tag{12}$$

$$Q_r^{(2)}(x) = \int^x \frac{\rho}{y} \Pi_r^{(2)}(y) dy = \frac{3}{4} \frac{x}{r} \left( 1 - \left( \frac{r}{\rho} \right)^2 \right) + \frac{3}{4} \frac{x^2}{\rho^2} - \frac{1}{4} \frac{x^3}{r \rho^2} \tag{13}$$

$$\rho = \omega/4 \tag{14}$$

## III. SIMULATION PROTOCOL

Starting from a random conformation obtained at $\sigma = 0$, a simulation run consists in starting with $\sigma = 0$ and repeating the following steps up to $\sigma = -0.08$:

1. Perform $\mathcal{N}$ sweeps at constant $\sigma$ (cf below for the parameters and quantities associated with the Monte-Carlo method)

2. Decrease $\sigma$ by 0.005

3. Goto 1

As a result, we have statistics for 17 values of $\sigma$ that are regularly spaced between $-0.08$ and $0$. The associated supercoiling rate, per sweep, of $\sigma$ variation is hence given by $v = -0.005/\mathcal{N}$, with $\mathcal{N} = 5 \times 10^5$ and $\mathcal{N} = 1.6 \times 10^7$ for the quickest and slowest simulations, respectively – in the following, for clarity, we normalize $v$ such that $v = 1$ for

4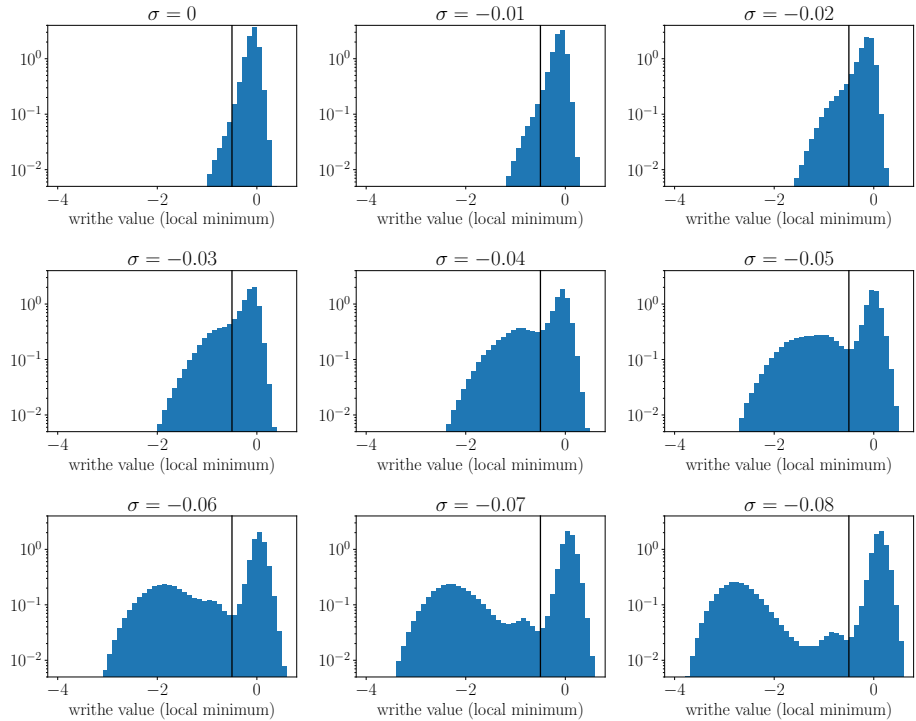

FIG. 2. *Distribution of local minima for the local writhe, wrs.* Each plot was obtained for a given value of $\sigma$. The vertical lines indicate the threshold wr$^*$ below which local minima are considered to be associated with plectonemic branches.

the quickest simulations (Fig. S1). Note that, in our simulations, a maximum of $M = 100$ cylinders can be rotated during a crankshaft rotation. Simulations being performed with a resolution of 30 bp per cylinder, a 30 kb long chain is made of $N = 1000$ cylinders such that a sweep corresponds to $N/M = 10$ Monte-Carlo steps. As a result, the slowest simulations with $\mathcal{N} = 1.6 \times 10^7$ corresponds to $\mathcal{N}_{MC} = 1.6 \times 10^8$ Monte-Carlo steps, so that one simulation run for the slowest case involves $17 \times 1.6 \times 10^8 = 2.72 \times 10^9$ Monte-Carlo steps.

For each simulation run, for further statistical analysis we have considered 2500 conformations between the $(N/2)^{th}$ sweep (mid-total number of sweeps) and the $N^{th}$ sweep (last sweep). For instance, Fig. S1 shows, for each supercoiling rate, the mean value of the radius of gyration as a function of $\sigma$ together with the standard error of the mean. The latter is computed using the variance of the mean of the radii of gyration obtained from the 20 different simulation runs, i.e. $\sqrt{\text{var}(R_g)/19}$ where var$(R_g)$ is that variance. Fig. S1 shows in particular that for rates smaller than $v = 1/8$, results may be considered independent of $v$. As a consequence, results of the main text, such as $P_s(r)$, have been obtained using 60 independent simulation runs coming from the three slowest rates ($v = 1/8, 1/16, 1/32$).

## IV. NUMBER OF PLECTONEMIC BRANCHES

The number of plectonemic branches (i.e. of external branches of the tree-like structures of supercoiled DNA) is computed using a local writhe, wr, as introduced in [1]. Specifically, for a given site $i$ of the chain ($i \in \{1..N\}$), $\text{wr}(i) = (2\pi)^{-1} \sum_{j=i-m/2}^{i+m/2} \Omega_{ij}$ where $\Omega_{ij}$ is given in Eqs. 16-21 in [2] (see [3] for the original derivation) and $m$ defines the window over wich the local writhe is computed. Here we take $m = 10$ such that it corresponds to two times the DNA bending persistence length – it is hence expected to be sensitive to the smallest plectonemes (i.e. curls). Next, for each conformation, we compute its profile of local writhes, that is, we compute the curve $(i, \text{wr}(i))$ with $i$ varying from 1 to $N$. We then identify the most significant local minima (largest negative values), which indicate *a priori* the presence of plectonemic branches. To this end, we compute distributions of all local minima over all studied conformations (Fig. 2). From these distributions, we define a threshold, wr$^*$ (vertical black lines in Fig. 2), that separates values associated with non-plectonemic DNA, on one hand, and values associated with plectonemes, on the other hand. Note that multimodal distributions are only present below $\sigma \simeq -0.04$, in accord with the observation that plectonemes do not manifest for too small supercoiling values.

Given wr$^*$, the number of plectonemic branches of a conformation is given by the number of local minima with wr $\leq$ wr$^*$.

---

## V. SUPPLEMENTARY FIGURES

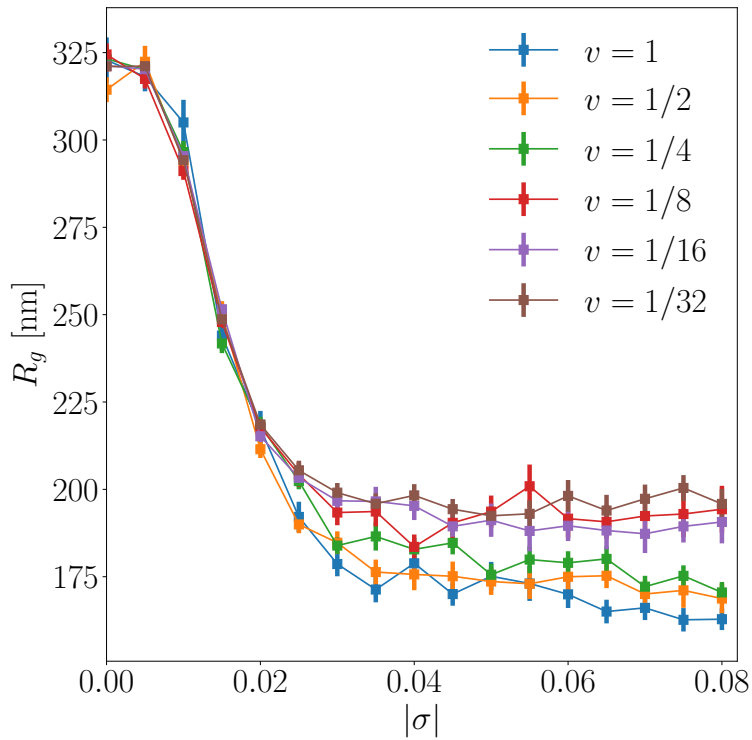

FIG. S1. *Sensitivity of results with respect to supercoiling rates.* In this plot, a point corresponds to the mean value of the radius of gyration obtained at a given $\sigma$ for a specific supercoiling rate $v$ (see explanations for the protocol). The error bars correspond to the standard error of the mean computed over 20 different simulation runs.

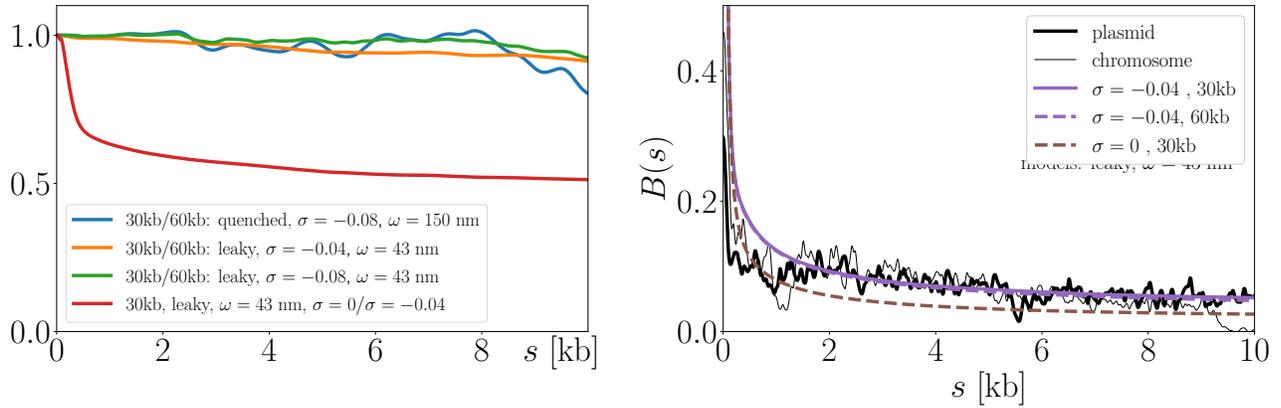

FIG. S2. *Below* 10 kb, *only slight differences exist between binding profiles obtained with* 30 kb *long molecules and those obtained with* 60 kb *long molecules.* **Left panel:** the blue, orange and green curves stand for the ratio of binding profiles between 30 kb and 60 kb long molecules obtained with different combinations of $\sigma$, $\omega$ and type of cluster. Red curve: ratio of binding profiles for a 30 kb long molecule with a leaky cluster and $\omega = 42$ nm between $\sigma = 0$ and $\sigma = -0.04$. **Right panel:** We report the binding profiles used to compute the orange and red curves on the left panel to demonstrate that differences between 30 kb and 60 kb long molecules are indeed not significant from the viewpoint of experimental data (the purple curves are indeed hardly distinguishable). By contrast, the difference is significative between $\sigma = 0$ (brown dashed curve) and $\sigma = -0.04$ (purple curves).

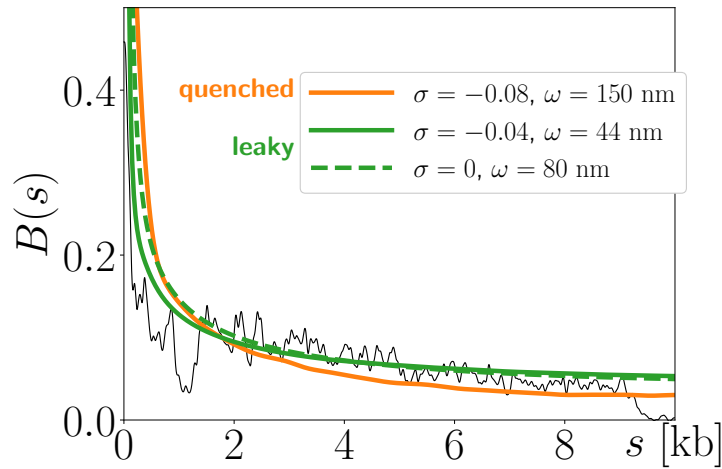

FIG. S3. *Capturing chromosomal binding profiles.* Black curve: ChIP-seq chromosomal data. Smooth plain curves: best models using a quenched cluster (in orange) or a leaky cluster (in green). Smooth dashed curve: best model at $\sigma = 0$ with a leaky cluster.



\end{document}